\documentclass[aps,prl,twocolumn,superscriptaddress,showpacs]{revtex4}

\usepackage{graphicx}
\usepackage{latexsym}
\usepackage{bm}
\usepackage{amsmath}

\usepackage{color}

\newcommand{\comment}[1]{}

\newcommand{\lsim}{\mbox{\raisebox{-0.6ex}{$\stackrel{<}{\sim}$}}\:}

\begin{document}

\title{
Pseudorapidity dependence of parton energy loss
    in relativistic heavy ion collisions
}

\author{Tetsufumi Hirano}
\affiliation{RIKEN BNL Research Center,
    Brookhaven National Laboratory, Upton, New York 11973}
\author{Yasushi Nara}
\affiliation{%
Department of Physics, University of Arizona, Tucson, Arizona 85721
}

\date{\today}
 
\begin{abstract}
We analyze the recent data from the BRAHMS Collaboration
on the pseudorapidity dependence of
nuclear modification factors in Au+Au collisions at $\sqrt{s_{NN}}$
= 200 GeV
 by using the full three dimensional
hydrodynamic simulations for the density
effects on parton energy loss.
We first compute the transverse spectra at $\eta=0$ and 2.2,
and next take a ratio $R_{\eta}=R_{AA}(\eta=2.2)/R_{AA}(\eta=0)$,
where $R_{AA}$ is a nuclear modification factor.
It is shown that hydrodynamic components account for $R_{\eta}\simeq 1$
 at low $p_\mathrm{T}$ and that quenched pQCD components lead $R_{\eta} < 1$
 at high $p_\mathrm{T}$ which are consistent with the data.
Strong suppression at $\eta=2.2$
is compatible with the parton energy loss in the final state.
\end{abstract}


\maketitle

Recent data from Relativistic Heavy Ion Collider (RHIC) reveal
  that hadron spectra
  at high $p_{\mathrm{T}}$ in central Au+Au collisions
  are strongly suppressed relative to
  the scaled $pp$ or large centrality spectra
  by the number of binary collisions~\cite{phenix:pi0_130,
phenix:pi0_200,star:highpt,star:charged,Adcox:2002pe}
contrary to the enhancement
in $d$+Au collisions~\cite{rhic:dA:phenix,rhic:dA:star,
rhic:dA:phobos,brahms:eta}.
The RHIC data is consistent with
  the early predictions on jet quenching
  due to gluon bremsstrahlung induced
  by multiple scattering~\cite{Gyulassy:1990ye}
  as a possible signal of deconfined nuclear matter,
  the quark gluon plasma (QGP)~\cite{Wang:2003aw}
  (for a recent review, see Ref.~\cite{Gyulassy:2003mc}).
Observed suppression of away-side peak in dihadron spectra
in central Au+Au collisions~\cite{star:btob}
is also considered to be due to jet quenching,
while
correlation spectra in $d$+Au collisions are the same
as in $pp$ collisions~\cite{rhic:dA:star},
  where it is not expected to create hot and dense matter.
Large elliptic flow observed in high-$p_{\mathrm{T}}$ region
 is also considered to be a consequence of jet quenching~\cite{Adler:2002ct,Adler:2003kt}.
Phenomenological studies
  based on the parton energy loss~\cite{Baier,Wiedemann,Zakharov,Levai}
  are successful in describing
 various high $p_\mathrm{T}$ hadron spectra at RHIC:
 suppression of single particle spectra~\cite{VG,jamal,HIRANONARA,Wang:btob},
 suppression of away-side correlation~\cite{HIRANONARA3,Wang:btob},
 azimuthal anisotropy of high-$p_\mathrm{T}$ hadron spectra in non-central collisions~\cite{Gyulassy:2000gk,Gyulassy:2001kr,Wang:btob,HIRANONARA4}
 including 
 centrality dependences~\cite{Wang:btob}.

In addition to those data, the BRAHMS Collaboration
recently reports
the pseudorapidity dependence of the nuclear
modification factors and shows that the yields of 
high $p_{\mathrm{T}}$ charged hadrons are
strongly suppressed even at $\eta=2.2$~\cite{brahms:eta}.
Furthermore, it is also shown that the ratio
  $R_{\eta}^{CP}=R_{CP}(\eta=2.2)/R_{CP}(\eta=0)$,
 where $R_{CP}$ is a ratio of
central to most peripheral yields normalized
 by the number of binary collisions, is
almost unity at $p_\mathrm{T}<2$ GeV/$c$
 and $R_{\eta}^{CP} < 1$ at high $p_\mathrm{T}$.
These data are the first results of high $p_{\mathrm{T}}$ spectra
in the forward rapidity region at RHIC and, thus, provide the novel
opportunity to study how the dense matter is distributed
in the longitudinal directions.
Therefore, further systematic studies are necessary to confirm 
the presence of the jet quenching in the dense medium at RHIC.
In this letter, we analyze $p_{\mathrm{T}}$ spectra at $\eta=2.2$
by employing the hydro+jet
model~\cite{HIRANONARA,HIRANONARA2,HIRANONARA3,HIRANONARA4}
and test if the scenario of jet quenching in the QGP phase 
 is still consistent with data.

Hydrodynamics
is found to be successful for the description of
the soft part of
the matter produced in Au+Au collisions
at RHIC especially in midrapidity region
($Y\simeq 0$)~\cite{Huovinen:2003fa}.
Motivated by these results,
we also describe the space-time evolution of thermalized matter
even in off-midrapidity region ($Y \ne 0$)
by solving the equations for energy-momentum conservation
in the \textit{full} three-dimensional (3D)
 Bjorken coordinate $(\tau,x,y,\eta_{\mathrm{s}})$~\cite{Hirano:2001eu,
HiranoTsuda}.
Here $\tau=\sqrt{t^2-z^2}$ is the proper time
 and $\eta_{\mathrm{s}}=(1/2)\ln[(t+z)/(t-z)]$
is the space-time rapidity.
Even at RHIC energies, one cannot observe
``central plateau~\cite{Bjorken:1982qr}" 
in the rapidity distribution~\cite{JHLEE} in Au+Au collisions.
Note that a plateau-like structure
in the \textit{pseudorapidity} distribution
observed at RHIC~\cite{Bearden:2001qq,Back:2001bq}
simply comes from the Jacobian between rapidity and pseudorapidity. 
In addition, elliptic flow as a
function of pseudorapidity shows a peak
at midrapidity~\cite{Back:2002gz}.
These data suggest
that the full 3D hydrodynamic 
simulations are necessary
for discussion on the global behavior in heavy ion collisions
even at RHIC energies.
Although our initial condition for longitudinal flow rapidity
(defined as $Y_{\mathrm{f}} = (1/2)\ln[(1+v_z)/(1-v_z)]$ where
$v_z$ is the longitudinal flow velocity)
is still assumed to be
the exact scaling solution
$Y_{\mathrm{f}}(\tau_0,x,y,\eta_{\mathrm{s}})=\eta_{\mathrm{s}}$,
 we obtain the rapidity dependent observables
by taking an
$\eta_{\mathrm{s}}$ dependent initial energy density distribution
which is factorized by a function
\begin{equation}
\label{eq:long}
H(\eta_{\mathrm{s}}) =
\exp\left[-\frac{(\mid \eta_{\mathrm{s}} \mid -\eta_{\mathrm{flat}}/2)^2}{2\eta_{\mathrm{Gauss}}^2} \theta(\mid \eta_{\mathrm{s}} \mid -\eta_{\mathrm{flat}}/2)\right],
\end{equation}
where $\eta_{\mathrm{flat}}$ and $\eta_{\mathrm{Gauss}}$ control the
size of a flat (Bjorken-like) region near midrapidity
and the width of Gaussian function in forward/backward rapidity respectively~\cite{Akase:1990yd}.
The pseudorapidity distributions
of charged hadrons in central and semi-central collisions
observed by BRAHMS~\cite{Bearden:2001qq}
are already reproduced by choosing $\eta_{\mathrm{flat}}/2 = 2$
and $\eta_{\mathrm{Gauss}}=0.8$ and also by
taking account of local rapidity shift
at each transverse coordinate $\eta_{\mathrm{s0}}(x,y)$
in Eq.~(\ref{eq:long})~\cite{HIRANONARA4}.
For details on how to parametrize initial conditions in our
hydrodynamic model,
see Ref.~\cite{HiranoTsuda}.
In hydrodynamic calculations,
 a partial chemical equilibrium (PCE) model
with chemical freeze-out temperature $T^{\mathrm{ch}}=170$ MeV
is employed for the hadronic phase
to describe the early chemical freeze-out picture
of hadronic matter~\cite{HiranoTsuda}. On the other hand,
the QGP phase
is assumed to
be massless free partonic gas with the number of flavor $N_f=3$.

%
\begin{figure}[t]
\includegraphics[width=3.3in]{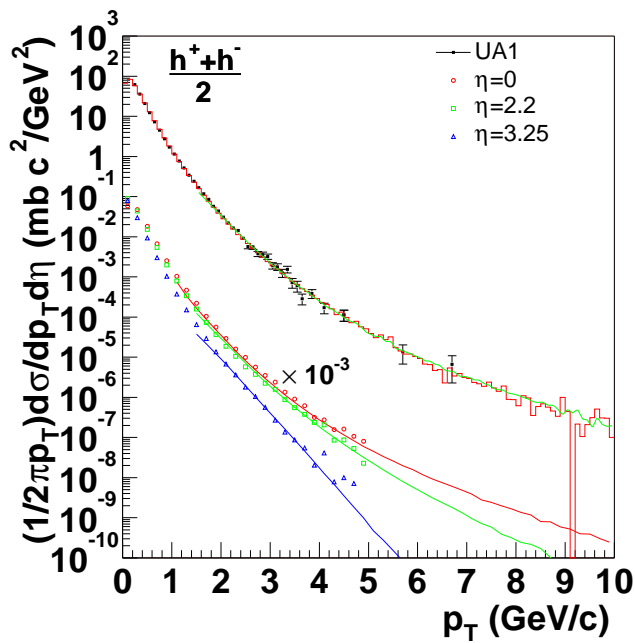}
\caption{
(Color online)
Invariant spectra of 
charged hadrons
from UA1 data~\cite{ua1}
in $p{\bar p}$ collisions at $\sqrt{s}=200$ GeV
is compared to PYTHIA predictions with Lund fragmentation (histogram)
and with independent fragmentation model (line).
$\eta$ dependence is also shown for both models.
Circles, squares, and triangles
correspond to the predictions from PYTHIA with Lund fragmentation
at $\eta=0,2.2$ and 3.25, respectively which are compared to
the results from PYTHIA with independent fragmentation model (lines).
}
\label{fig:dndptUA1}
\end{figure}
%

For the hard part of the model,
  we generate hard partons according to a pQCD parton model.
 We use PYTHIA 6.2~\cite{pythia} for the generation
of momentum spectrum of jets through
  $2\to2$ QCD hard processes.
  Initial and final state radiations are used
  to take into account the enhancement of higher-order contributions
  associated with multiple small-angle parton emission.
The CTEQ5 leading order parton distribution function~\cite{cteq5} is used.
Hadrons are obtained from
an independent fragmentation model option in PYTHIA.
The $K$-factor $K=2.5$,
the scale $Q=p_{\mathrm{T,jet}}/2$ in the parton distribution function,
  and the primordial transverse momentum
  $\langle k_{\mathrm{T}}^2\rangle_{NN}=1.2$ GeV$^2$/$c^2$
  are used to fit 
  the neutral pion transverse spectrum in $pp$ collisions
  at RHIC~\cite{Adler:2003pb}.
UA1 data~\cite{ua1} for charged hadrons are also well reproduced with
this parameters in the transverse momentum range $p_\mathrm{T}>2$ GeV/$c$
as shown in Fig.~\ref{fig:dndptUA1}.
Independent fragmentation model is not applicable for the low
$p_\mathrm{T}$ transverse momentum range.
Hence we use the Lund string model for the $pp$ reference
when we compute nuclear modification factors
at low $p_\mathrm{T}$ ($p_\mathrm{T}< 2$ GeV/$c$).
EKS98 parametrization~\cite{eks98}
is employed to take into account nuclear shadowing effect.
The Cronin enhancement is modeled by 
 the multiple initial state scatterings as in Ref.~\cite{XNWang:1998ww}.

Initial transverse positions of jets
 are determined randomly
 according to 
 the number of binary collision distribution.
Initial longitudinal position of a parton
 is approximated by the boost invariant distribution~\cite{Bjorken:1982qr}.
Jets are freely propagated up to the initial time $\tau_0$(=0.6 fm/$c$)
 of hydrodynamic
simulations
by neglecting the possible interactions in the pre-thermalization stages.
Jets are assumed to travel with straight line trajectory
in the medium.

Jets can suffer interaction with fluids and lose their energies.
We employ the approximate first order formula (GLV formula)
 in opacity expansion
 from the reaction operator approach~\cite{Levai}
 for the energy loss of partons in this work.
The approximate first order formula in this approach can be written as
\begin{equation}
\Delta E = C \int_{\tau_0}^{\infty} d\tau
\rho\left(\tau, \bm{x}\left(\tau\right)\right)
(\tau-\tau_0)\ln\left({\frac{2E_0}{\mu^2 L}}\right).
\label{eq:GLV}
\end{equation}
Here we take $L=3$ fm and $\mu=0.5$ GeV.
$C=0.45$ is an adjustable parameter and
$\rho(\tau,\bm{x})$ is a thermalized parton density
 in the local rest frame of fluid elements in the hydro+jet approach
\cite{noteweb}.
$\bm{x}\left(\tau\right)$ and $E_0$ are the position and
the initial energy of a jet, respectively.
The initial energy $E_0$ in Eq.~(\ref{eq:GLV}) 
is Lorentz-boosted by the flow velocity and replaced by $p_0^{\mu} u_{\mu}$
where $p_0^\mu$ and $u_{\mu}$ are the initial four momentum of a jet and a local fluid velocity respectively.
The parameters related to the propagation
of partons are obtained by fitting
 the nuclear modification factor for
  the neutral pion by PHENIX~\cite{phenix:pi0_200}
  and are found to be 
  consistent~\cite{HIRANONARA3}
   with the back-to-back correlation data from STAR~\cite{star:btob}.


%
\begin{figure}[t]
\includegraphics[width=3.3in]{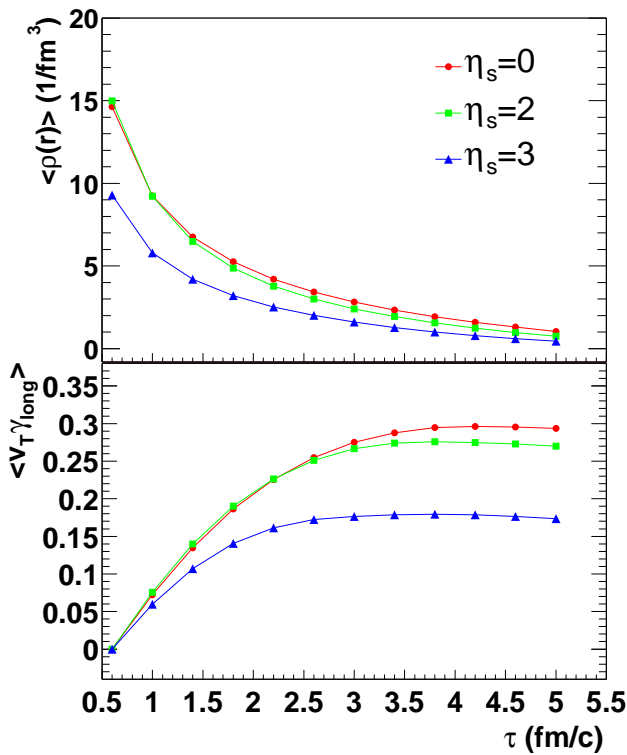}
\caption{
(Color online)
Time evolutions of average parton density (upper panel)
and average $\gamma_{\mathrm{long}}$-weighted
transverse velocity (lower panel)
at space-time rapidity $\eta_s=0$, 2, and 3
from hydrodynamic simulations
in 0-10\% central Au+Au collisions
at $\sqrt{s_{NN}}=200$ GeV.
Impact parameter is $b=3.7$ fm.
Average is taken only in the QGP and mixed phases.
}
\label{fig:rho_vt_b37}
\end{figure}
\begin{figure}[t]
\includegraphics[width=3.3in]{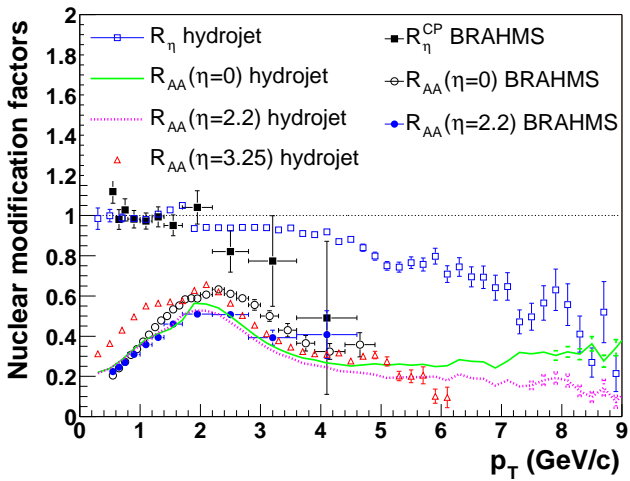}
\caption{
(Color online)
Nuclear modification factors are compared to
the BRAHMS data~\cite{brahms:eta}
in Au+Au collisions at $\sqrt{s_{NN}}=200$ GeV.
}
\label{fig:Reta}
\end{figure}

%
Let us start with the study of
 the space-time rapidity dependence of transverse dynamics
from the full 3D hydrodynamics.
In order to understand the dynamical effects on parton energy loss,
we plot in Fig.~\ref{fig:rho_vt_b37}
parton densities and 
$\gamma_{\mathrm{long}}$-weighted transverse flow velocities
averaged over the QGP and mixed phases
as a function of $\tau$ at $\eta_{s}=0, 2$, and 3.
Here a gamma factor of longitudinal flow is
$\gamma_{\mathrm{long}} = \cosh(Y_{\mathrm{f}})$.
Note that the quantity
 $\langle v_{\mathrm{T}} \gamma_{\mathrm{long}} \rangle$
is independent of the space-time rapidity $\eta_{\mathrm{s}}$
when the initial longitudinal flow and the initial energy density
 obey the scaling solution
$Y_{\mathrm{f}}(\tau_0,x,y,\eta_{\mathrm{s}})=\eta_{\mathrm{s}}$ and
$e(\tau_0,x,y,\eta_{\mathrm{s}}) = e(\tau_0,x,y)$ respectively.
We note that longitudinal flow from full 3D hydrodynamics
remains to be close to the Bjorken flow:
 $Y_f \approx \eta_{\mathrm{s}}$
 at $\eta_{\mathrm{s}}<3$
 within our initial conditions used in this work.
As shown in the figure,
the difference of the time evolutions of
  the thermalized parton density and the transverse flow velocity
  between $\eta_{\mathrm{s}}=0$ and $\eta_{\mathrm{s}}=2$
  are practically tiny,
since the initial scaling region reaches to
$\eta_{\mathrm{s}} = \eta_{\mathrm{flat}}/2 = 2$~\cite{HIRANONARA5}.
Thus, the dynamical effect on the jet energy loss
at $\eta=0$ and 2.2 is expected to be the same,
  whereas the amount of jet energy loss at $\eta=3$ 
  must be small because of the smaller parton density.
As will be shown in Fig.~\ref{fig:Reta},
we should emphasize here that this does not mean
$R_{AA}(\eta=0) \approx R_{AA}(\eta=2.2)$.
\begin{figure}[t]
\includegraphics[width=3.3in]{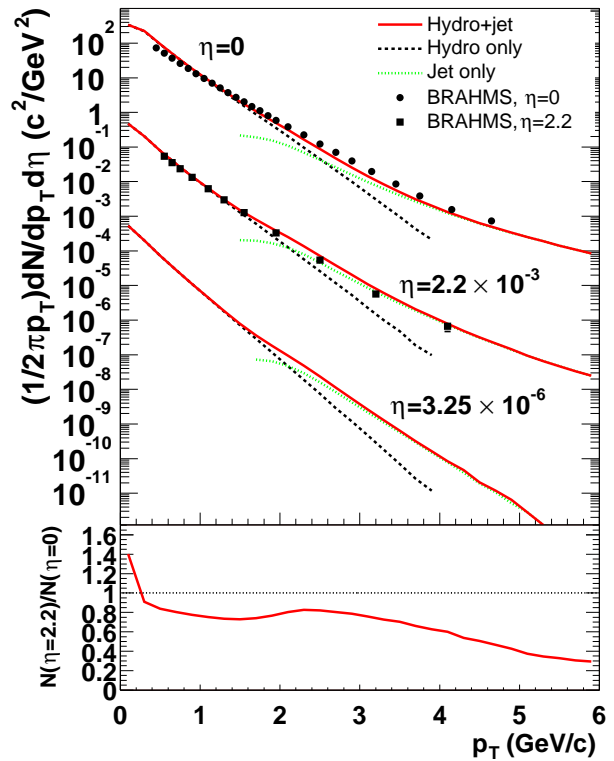}
\caption{
(Color online)
Transverse momentum distributions of charged hadrons
 in Au + Au  collisions at $\sqrt{s_{NN}}=200$ GeV
is compared to data from BRAHMS~\cite{brahms:eta}.
Solid lines represent the hydro+jet result
averaged over $|\eta|<0.1$,  $2.1<|\eta|<2.3$, and $3.0<|\eta|<3.5$.
The impact parameter for 0-10\% centrality is taken to be
$b=3.7$ fm.
In the bottom panel, the ratio of the spectrum at $\eta=2.2$
to that of $\eta=0$ is plotted.
Enhancement at very low $p_\mathrm{T}$ comes from the effect of Jacobian
for the transformation of rapidity to pseudorapidity.
}
\label{fig:dndpt_hydrojet}
\end{figure}

We next show in Fig.~\ref{fig:dndpt_hydrojet}
the transverse momentum distributions
 for charged hadrons
 from the hydro+jet model
 in central Au+Au collisions at RHIC.
Thermal freeze-out temperature $T^{\mathrm{th}}=100$ MeV
  is used in the calculation.
This choice is consistent with the PHENIX data
  at $\sqrt{s_{NN}}=130$ GeV~\cite{HiranoTsuda}.
Each spectrum is the sum of the soft component and the hard component.
Before summation, the hard component is multiplied
by a ``switch" function~\cite{Gyulassy:2000gk}
$\{1+\tanh[2(p_{\mathrm{T}}-2)]\}/2$
(where $p_{\mathrm{T}}$ is in the unit of GeV/$c$)
in order to cut the unreliable components
from the independent fragmentation scheme
and also to obtain the smooth spectra.
The hydrodynamic components are dominated
in the range of $p_\mathrm{T}<2$ GeV/$c$.
The slope of hadrons in low $p_{\mathrm{T}}$ region
at $\eta=0$
is nearly the same as the one at $\eta=2.2$
as clearly seen in the bottom panel of Fig.~\ref{fig:dndpt_hydrojet} in which the ratio of the spectrum at $\eta=2.2$ to that at $\eta=0$ is plotted.
This results from the similarity of transverse dynamics
between $\eta=0$ and 2 as shown in Fig.~\ref{fig:rho_vt_b37}.
On the other hand,
the slope from pQCD components in high $p_{\mathrm{T}}$ region
becomes steeper as $\eta$ increases
reflecting the original $pp$ spectra (see Fig.~\ref{fig:dndptUA1}).

We now turn to the study of the nuclear modification factors $R_{AA}$
 for charged hadrons
 defined by
\begin{equation}
\label{eq:raa}
 R_{AA} = \frac{\frac{dN^{A+A}}{d^2p_{\mathrm{T}}d\eta}}
    {N_{\mathrm{coll}}\frac{dN^{p+p}}{d^2p_{\mathrm{T}}d\eta}},
\end{equation}
where $N_{\mathrm{coll}}$ is a number of binary collisions.
Figure~\ref{fig:Reta} shows the nuclear modification factors $R_{AA}$ for
   charged hadrons at $\eta=0$, 2.2, and 3.25
   in Au+Au collisions at RHIC
   for an impact parameter $b=3.7$ fm.
The nuclear modification factor $R_{AA}$'s
 in low $p_\mathrm{T}$ region ($p_\mathrm{T}\lsim 2$ GeV/$c$),
 where the hydrodynamic component dominates,
 at $\eta=0$ and 2.2 are almost identical.
This is due to the comparable time evolution of
 the parton density at $\eta=0$ and 2.2
 in hydrodynamics as shown in Fig.~\ref{fig:rho_vt_b37}.
PYTHIA prediction reveals that the spectrum at low $p_{\mathrm{T}}$ is
 very similar within $\eta<2$ up to a factor of 30\%
 in $pp$ collisions as shown in Fig.~\ref{fig:dndptUA1}.
$R_{AA}(\eta=0) > R_{AA}(\eta=2.2)$ at high $p_{\mathrm{T}}$
 is a consequence of the steeper slope at $\eta=2.2$
 compared to the slope at $\eta=0$ in pQCD.
When the $p_{\mathrm{T}}$ slope is steep, the nuclear modification
factor becomes sensitive to nuclear effects:
a small shift of a spectrum is likely to
produce a large effect on the ratio of the shifted spectrum
to the original one.
It should be noted
that, due to the above reason, the nuclear modification factor from 
the Cronin effect at SPS energies~\cite{Aggarwal:2001gn} is
much larger than the one at RHIC energies~\cite{rhic:dA:phenix,rhic:dA:star,rhic:dA:phobos,brahms:eta}.
The nuclear modification factor at $\eta=3.25$
in the range $p_{\mathrm{T}}<5$ GeV/$c$
 is larger than at midrapidity,
 because thermalized parton density at $\eta=3.25$
is about 40\% smaller than at midrapidity.
However, $R_{AA}(\eta=3.25)$
eventually becomes smaller than the one for $\eta=0$ or 2.2
in high $p_{\mathrm{T}}$ region.
This is due to the much steeper slope at high $p_{\mathrm{T}}$.

We also plot the ratio $R_{\eta}$ defined by
\begin{equation}
 R_{\eta}= \frac{R_{AA}(\eta=2.2)}{R_{AA}(\eta=0.0)},
\end{equation}
which can be compared with the data from BRAHMS
defined by
$R^{CP}_{\eta}=R_{CP}(\eta=2.2)/R_{CP}(\eta=0.0)$.
Hydrodynamic predictions at very peripheral collisions are likely to be
unreliable, since parton density is too small to justify
the assumption of the local thermal system
in Au+Au collisions at RHIC.
In Fig.~\ref{fig:Reta}, $R_\eta$ is found to be almost unity
in low $p_{\mathrm{T}}$ region and 
gradually decreases with $p_{\mathrm{T}}$ as consistent with data
as expected in the above consideration.

One may worry about the nuclear shadowing effect
on the $R_{\eta}$,
since
the accessible range of the parton momentum fraction $x$
 in the parton distribution functions at $\eta=2.2$
 is smaller than at midrapidity.
However,
nuclear shadowing effect on $R_{AA}(\eta=2.2) < R_{AA}(\eta=0)$
 is found to be $\sim 15$ \%
around $p_{\mathrm{T}} = 2$-3 GeV/$c$  within EKS98 parameterization.
Other models for nuclear shadowing should be investigated elsewhere.

\begin{figure}[t]
\includegraphics[width=3.3in]{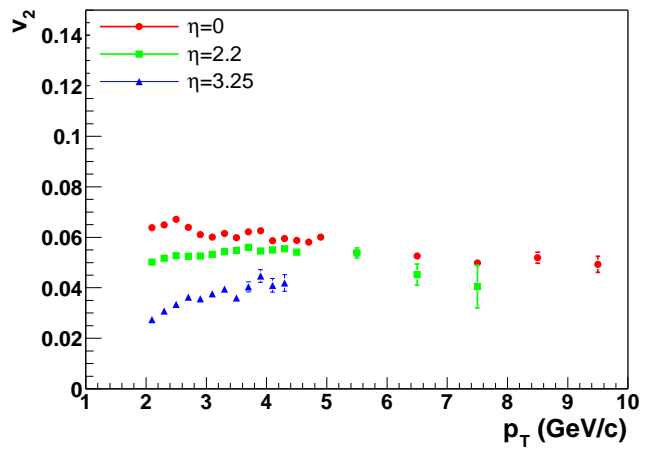}
\caption{
(Color online)
Elliptic flow parameter $v_2$ for charged hadrons
at $\eta=0$, 2.2, and 3.25.
The impact parameter is taken as $b=7.2$ fm.
 We only take into account
contributions from hard components.
}
\label{fig:v2}
\end{figure}

In order to disentangle the density effect from the ``slope"
effect among nuclear modification factors, we predict the
elliptic flow parameter $v_2=\langle \cos (2\phi)\rangle$
for high $p_{\mathrm{T}}$
charged hadrons in forward rapidity region as shown in
Fig.~\ref{fig:v2}.
In this calculation, we only take account of the
contributions from hard components. We choose the impact
parameter as $b=7.2$ fm corresponding to 20-30\% centrality.
$v_2$ in high $p_{\mathrm{T}}$ region is generated by jet quenching
\cite{Gyulassy:2000gk}. The difference of path length
causes the difference of the amount of parton energy loss
in azimuthal directions in non-central collisions.
Consequently, one can expect from Eq.~(\ref{eq:GLV}) that the higher
parton density results in the larger positive $v_2$
until $v_2$ reaches the limiting value~\cite{Shuryak:2001me}.
As we expected, our prediction $v_2(\eta=3.25) < v_2(\eta=0)$
 is a clear evidence of the density effect on the parton energy loss.
We comment on our prediction on the $v_2$ at high $p_{\mathrm{T}}>5$
GeV/$c$
at midrapidity. Our prediction is similar to the results of the recent
work \cite{Drees:2003zh} in which they found that
the azimuthal anisotropy of high $p_T$ particles underestimate
for a realistic nuclear density profile although hard-sphere nuclear
profile looks consistent with the data~\cite{Wang:btob}.

In summary,
we have studied the pseudorapidity dependence of the
nuclear modification factors
for charged hadrons
within the hydro+jet model.
As well as the yield of charged hadrons,
the radial flow at $\eta_{\mathrm{s}}=2$ is found to be
very similar to that at $\eta_{\mathrm{s}}=0$.
This results in $R_{\eta}\simeq 1$ at $p_\mathrm{T} < 2$ GeV/$c$.
$R_{\eta} < 1$ in high $p_\mathrm{T}$ region can be understood
by the steeper slope of $p_{\mathrm{T}}$ spectrum
at $\eta=2.2$ than at $\eta=0$
from the pQCD components.
This suggests that the longitudinal region of 
dense partonic matter
produced in Au+Au collisions
reaches to $\eta_{\mathrm{s}} \sim 2$
and that 
strong hadron suppression at off-midrapidity is consistent with
the final state parton energy loss in the medium.
This reminds us the previous
analysis of pseudorapidity dependence
of elliptic flow at the RHIC energy~\cite{Hirano:2001eu,HiranoTsuda}
in which elliptic flow can be reproduced by hydrodynamics
only in the region $\mid \eta \mid \lsim 2$. 
We also predicted the elliptic flow parameter $v_2$
for high $p_{\mathrm{T}}$ charged hadrons in both midrapidity and
forward rapidity regions.
We found the strong rapidity dependence of $v_2 (p_{\mathrm{T}})$.

It would be very interesting to see how robust the present
results are in more realistic analysis.
For example, the lattice studies predict
the non-trivial behavior of screening mass in the
vicinity of phase transition region~\cite{Kaczmarek:1999mm}.
This affects the space-time evolution of the parton
density in
the energy loss formula in Eq.~(\ref{eq:GLV}).
The parton density is multiplied by the correction
factor for the effective degree of freedom $\mu^2/\mu_0^2$
\cite{Pisarski:2002ji}
where $\mu_0$ and $\mu$ are, respectively, perturbative
and non-perturbative screening mass.
The reduction of energy loss due to the non-perturbative
behavior of the screening mass
is expected to be large
in forward rapidity region where
the initial parton density is not so large.
This modification causes the change of equation of state (EOS)
as well as the amount of the parton energy loss.
 Hence we may need to use different
initial conditions and simulate the hydrodynamic model
with a more realistic EOS.
This is beyond the scope of this paper
although it should be revisited elsewhere.

\begin{acknowledgements}
The work of T.H.~is supported by 
RIKEN.
Y.N.'s research is supported by the DOE under Contract No.
DE-FG03-93ER40792.
\end{acknowledgements}

\end{document}